# A mode-multiplexed photonic integrated vector dot-product core from inverse design

Zheyuan Zhu,[1,][*] Raktim Sarma,[2] Seth Smith-Dryden,[1] Guifang Li,[1] and Shuo S. Pang[1]

[1]*CREOL, The College of Optics and Photonics, University of Central Florida, 4304 Scorpius St. Orlando, FL 32816-2700, USA*
[2]*Center for Integrated Nanotechnologies, Sandia National Laboratories, 1515 Eubank Blvd SE, Albuquerque, NM 87185, USA*
*[*]zyzhu@knights.ucf.edu*

**Abstract:** Photonic computing has the potential of harnessing the full degrees of freedom (DOFs) of the light field, including wavelength, spatial mode, spatial location, phase quadrature, and polarization, to achieve a higher level of computing parallelism and scalability than digital electronic processors. While multiplexing using wavelength and other DOFs can be readily integrated on silicon photonics platforms with compact footprints, conventional mode-division multiplexed (MDM) photonic designs occupy areas exceeding tens to hundreds of microns for a few spatial modes, significantly limiting their scalability. Here we utilize inverse design to demonstrate an ultracompact photonic computing core that calculates vector dot-products based on MDM coherent mixing. Our dot-product core integrates the functionalities of 2 mode multiplexers and 1 multi-mode coherent mixers within a nominal footprint of 5 µm × 3 µm. We have experimentally demonstrated computing examples on the fabricated dot-product core, including complex number multiplication and motion estimation using optical flow. The compact dot-product core design enables large-scale on-chip integration in a parallel photonic computing primitive cluster for high-throughput scientific computing and computer vision tasks.

## 1. Introduction

Vector, matrix, or tensor calculations are fundamental building blocks of modern scientific computing. The underlying core components of these computing tasks are basic linear algebra subprograms (BLAS) that provide hardware implementation of the arithmetic operations between vectors (level 1), vector and matrix (level 2), and matrices (level 3), each building upon the previous level [1]. Given its unique role as a BLAS level 1 routine, efficient and scalable vector dot-product calculation is crucial to achieving optimal performances in more complex and computationally intensive operations. In traditional uniprocessor digital computers, the central processing unit (CPU) executes a single basic operation, such as addition, multiplication, or fused multiply-add (FMA), on a single data stream, a process known as single instruction stream, single data stream (SISD, Figure 1(a)). The sequential execution and repeated data access of SISD compromise the computation speed and efficiency in vector- and matrix-based operations. Single instruction stream, multiple data streams (SIMD, Figure 1(b)), which simultaneously applies an arithmetic operation to multiple data streams [2], has been adopted in virtually all modern CPUs and stream processors in GPUs. These processors incorporate dedicated SIMD tiles of cascaded FMA units with pipelined inputs to accelerate vector instructions [3]. Because caching the intermediate results is still necessary to ensure timing closure in electronics, the computing throughput per unit area is usually on the order of 0.1 tera operations per second per millimeter square (TOPS/mm$^2$), and the vector length is typically limited to several hundreds, even with highly optimized layout of logic and memory units within an SIMD engine [3,4].

Recently, driven by the computing demand in the artificial intelligence (AI) industry, analog computing platforms based on integrated photonic devices [5,6] have demonstrated the potential of higher efficiency and computing throughput than the electronic counterparts, due to the intrinsically passive photonic multiply-accumulate (MAC) operations without intermediate memory access [7]. Figure 1(c) illustrates a photonic computing design based on two coherent mixers without parallelization in DOF of light, much like the SISD architecture in digital computing. In a single coherent mixing unit, the two inputs of electrical fields encode the numbers $a$ and $b$ in their amplitudes. After splitting and balanced detection, the output is proportional to their product $\mathrm{Re}\{a^*b\}$ [8]. To perform dot-products between two $N$-element vectors, $N$ sets of mixers and balanced photodiodes are required, and the intermediate element-wise products must first be individually digitized and then summed in the post processing stage. Due to the power consumption of analog-to-digital converters (ADCs) [9] required in the design, coherent mixing without data-level parallelism suffers from low efficiency when handling large vectors.

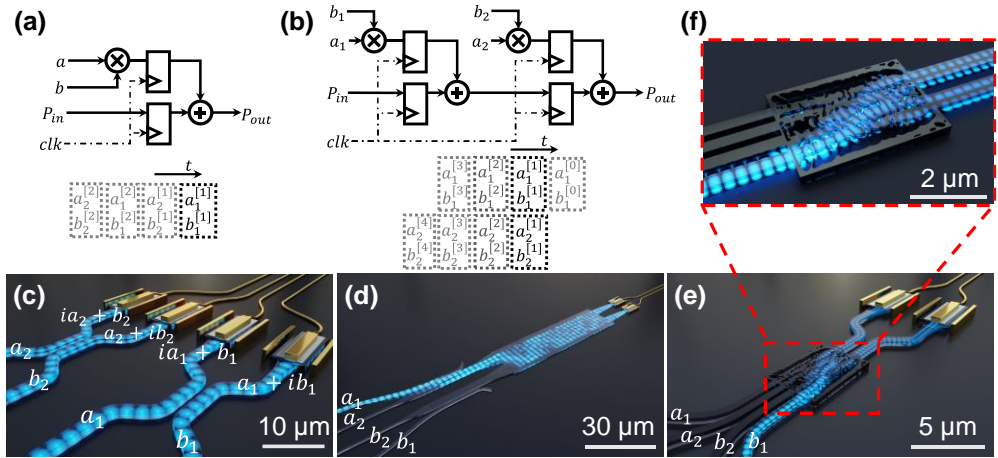

Figure 1. Comparison between electronic (a-b) and photonic (c-e) implementations of vector dot-product cores: (a) An SISD electronic arithmetic unit that performs multiplication and addition. (b) SIMD design with multiple, pipelined inputs for dot-product calculation. (c) Individual single-mode coherent mixers as multiplier units without parallelism, equivalent to SISD architecture in digital electronics; (d, e) Mode-multiplexed implementations of dot-product core in photonics, equivalent to SIMD architecture in digital electronics. (d) is based on based on the current MDM infrastructure in optical communications, using two MUX and one MMI; (e) is an end-to-end photonic dot-product core that integrates the functionalities of two MUX and one MMI. The inset (f) shows the photonic structure from inverse design.

Similar to the transition from SISD to SIMD architecture in digital processors, using wavelength- or mode-division multiplexed (WDM or MDM) photonic signals enables a single coherent detection unit, consisting of a 2×2 coherent mixer and a pair of balanced photodiodes, to simultaneously process multiple data inputs in parallel [10]. Leveraging the intrinsic orthogonality of the light fields, coherent photonic MAC operations with multiplexed signals naturally accumulates the intermediate elementwise products between two vectors, and thus could achieve 2-3 folds lower power consumption than the non-multiplexed designs [11]. While WDM-based photonic processing devices have matured into practice to some extent in AI-related computing applications [12,13], MDM-based devices only begin to emerge as a viable approach in high-bandwidth optical communication [14,15], and their applications in parallel photonic computing are yet to be exploited.

While it is clear that utilizing MDM can lead to significant advances in high-bandwidth optical communication and photonic computing, a major bottleneck to high density integration is the large footprint usually associated with these MDM-based nanophotonic devices. Figure 1(d) shows an implementation of photonic dot-product core based on conventional MDM components in optical communications. The elements in the vector, $a_1$ ($b_1$) and $a_2$ ($b_2$), are mapped to the fundamental ($\psi_\mathrm{I}$) and the second order ($\psi_\mathrm{II}$) TE modes of a few-mode waveguide via mode multiplexers (MUXs). The mode-multiplexed photonic signals, $E_a = a_1\psi_\mathrm{I} + a_2\psi_\mathrm{II}$ and $E_b = b_1\psi_\mathrm{I} + b_2\psi_\mathrm{II}$, undergo coherent mixing via multi-mode interference (MMI), producing the electrical fields on the upper and lower arms $E_p = \frac{1}{\sqrt{2}}(E_a + iE_b)$ and $E_n = \frac{1}{\sqrt{2}}(iE_a + E_b)$. Based on the orthogonality between $\psi_\mathrm{I}$ and $\psi_\mathrm{II}$, the difference between the overall intensity of the upper and lower outputs $I_{diff} = |E_p|^2 - |E_n|^2$ produces the dot-product between vectors $\vec{a}$ and $\vec{b}$. The functionality of the conventional MDM dot-product design can be expressed as a Kronecker product (denoted by $\otimes$) between a 3-dB coupling matrix representing the MMI, and an identity matrix representing the MUX, as in Eq. (1).

$$\begin{bmatrix} E_{\mathrm{II}p} \\ E_{\mathrm{I}p} \\ E_{\mathrm{I}n} \\ E_{\mathrm{II}n} \end{bmatrix} = \frac{1}{\sqrt{2}} \begin{bmatrix} 1 & i \\ i & 1 \end{bmatrix} \otimes \begin{bmatrix} 1 & 0 \\ 0 & 1 \end{bmatrix} \begin{bmatrix} a_1 \\ a_2 \\ b_2 \\ b_1 \end{bmatrix}. \tag{1}$$

Using conventional MDM components, the dot-product core requires 2 MUXs and 1 MMI. Each MUX occupies at least 20µm×4µm in footprint [16,17], which is required by the adiabatic taper. For a 2×2 MMI, a footprint of 40µm×6µm [18,19] is required to match the first Talbot distance. The overall footprint of a conventional dot-product core is thus larger than 50µm×10µm.

In this work, we present a topologically optimized mode-multiplexed photonic vector dot-product core (Figure 1(e)) that integrates the functionalities of 2 MUXs and 1 MMI within a 5 µm × 3µm footprint. Compared with the behavior of the electrical field inside a conventional multi-mode photonic design (Figure 1(d)), in which the regions for mode conversion and mixing are clearly distinguishable, the integrated dot-product core does not perform an intermediate conversion step of the input electrical fields onto the spatial mode basis. The end-to-end transformation of the electrical field by the integrated core, expressed as a matrix $S_i$ in Eq. (2), contributes to its compact footprint.

$$\begin{bmatrix} E_{\mathrm{II}p} \\ E_{\mathrm{I}p} \\ E_{\mathrm{I}n} \\ E_{\mathrm{II}n} \end{bmatrix} = \frac{1}{\sqrt{2}} \begin{bmatrix} 1 & 0 & 0 & i \\ 0 & 1 & i & 0 \\ 0 & i & 1 & 0 \\ i & 0 & 0 & 1 \end{bmatrix} \begin{bmatrix} a_1 \\ a_2 \\ b_2 \\ b_1 \end{bmatrix}. \tag{2}$$

The ultra-compact footprint addresses one of the fundamental bottlenecks for utilizing MDM-based approaches for photonic computing and paves the way for high-density integration of the core in a parallel computing array.

## 2. Methods

The photonic core was inversely designed on a silicon-on-insulator (SOI) platform by optimizing the structure that maximizes the coupling efficiency from the inputs into the target electric field profiles. The design process follows a gradient-based paradigm that tunes the distribution of the relative permittivity, $\boldsymbol{\varepsilon}_r$, on the silicon layer as the design parameters [26]–[29]. The parameters are updated along the gradient direction of the objective function (Eq. (3)),

$$l = \sum_{j=1}^{J} \left| \mathbf{E}_{t_j}^\dagger \mathbf{E}_j(\boldsymbol{\varepsilon}_r) \right|^2. \tag{3}$$

Here $l$ calculates the overlap integral between the target field $\mathbf{E}_{t_j}$ at the output location and the field $\mathbf{E}_j$ within the structure, $\boldsymbol{\varepsilon}_r$ is the three-dimension distribution of relative permittivity, and $(\cdot)^\dagger$ denotes the matrix conjugate transpose. We set the fundamental or second-order TE eigenmodes in the few-mode output waveguides as the target fields $\mathbf{E}_{t_j}$. The summation over $j$ aggregates the contributions from all four pairs of output and target fields. The field $\mathbf{E}_j$ in the device satisfies the finite-difference frequency domain (FDFD) Maxwell equations in matrix form,

$$\mathbf{A}_{PML}(\boldsymbol{\varepsilon}_r)\mathbf{E}_j(\boldsymbol{\varepsilon}_r) = \mathbf{b}_j. \tag{4}$$

Here $\mathbf{A}_{PML}(\boldsymbol{\varepsilon}_r)$ is the wave matrix [30] for three-dimensional vector electrical filed, representing the operator $\frac{1}{k_0}\nabla \times \nabla \times - \boldsymbol{\varepsilon}_r \odot$ with perfectly matched layers (PMLs) on the boundary of the solution domain [31], where $\odot$ denotes elementwise product, and $k_0$ is the wavenumber in vacuum. $\mathbf{b}_j$ is the input excitation that induces the field $\mathbf{E}_j$ within the device.

Combining Eq. (3) and (4), and noting that $\boldsymbol{\varepsilon}_r$ is separable from the finite difference operators in the wave matrix, the gradient of the objective function with respect to $\boldsymbol{\varepsilon}_r$ can be derived as

$$\nabla l(\boldsymbol{\varepsilon}_r) = \sum_{j=1}^{J} \text{Re}\{diag(\mathbf{E}_j^*)\mathbf{A}_{PML}^{-1}(\boldsymbol{\varepsilon}_r)\mathbf{E}_{tj}\}. \tag{5}$$

Here $\odot$ denotes the element-wise product. The inverse problem $\mathbf{A}_{PML}^{-1}(\boldsymbol{\varepsilon}_r)\mathbf{E}_{tj}$ was solved using the least squares method with MATLAB's LSQR function [32] and carried out on $J = 4$ parallel GPUs (NVIDIA RTX 3090). The relative permittivity $\boldsymbol{\varepsilon}_r$ is updated along the gradient direction with an adaptive step size $\tau$ as $\boldsymbol{\varepsilon}_r \leftarrow \boldsymbol{\varepsilon}_r + \tau \nabla l(\boldsymbol{\varepsilon}_r)$. To promote binary medium (air and silicon) on the silicon layer, the updated $\boldsymbol{\varepsilon}_r$ is mapped by a sigmoid function to produce the relative permittivity in the next iteration,

$$\boldsymbol{\varepsilon}_r' = \frac{\varepsilon_{Si} - \varepsilon_{air}}{1 + \exp\left(-\gamma\left(\boldsymbol{\varepsilon}_r - \frac{\varepsilon_{air} + \varepsilon_{Si}}{2}\right)\right)} + \varepsilon_{air}. \tag{6}$$

Here $\varepsilon_{Si}$ and $\varepsilon_{air}$ are the relative permittivity of silicon and air, respectively, and $\gamma = 4$ is a hyperparameter that controls the slope of the sigmoid function.

The inversely-design photonic dot-product core was fabricated on commercially available silicon-on-insulator (SOI) wafers. The wafers consisted of 250 nm silicon on top of a 3 µm buried oxide. The core was fabricated using a positive tone ZEP resist followed by electron beam lithography and inductively coupled plasma reactive ion etching. To realize the subwavelength sized and spaced features of the inversely designed structure, short range proximity correction was used to vary the dose of the exposure across the device. The core consisted of 4 single-mode input waveguides (480 nm in width) and 2 few-mode output waveguides (774 nm in width). The input waveguides were separated by 127 µm to match the pitch of an edge-coupling fiber array (Precision Micro-Optics Inc., FPFA-10208111212). The 2 few-mode output waveguides were each tapered to a 40 µm × 40 µm photonic crystal structure [33], which vertically couples out the electric field profiles for observation by a camera.

The four elements in the two input vectors were generated from four off-chip fiber-based Mach-Zehnder modulators (MZM, JDSU, OC-192). The modulators were driven by a multi-channel digital-to-analog converter (DAC, Analog Devices, MAX11300), which was controlled by a microcontroller (Analog Devices, SDP-CK1Z). The modulated signals were edge-coupled into the four input ports of the dot-product core. The intensity profiles on the two vertical output couplers were recorded from above through a long-working distance 20X objective and a short-wave infrared (SWIR) camera (Allied Vision, Goldeye CL-008 TEC1). The camera and DAC synchronously perform 100 multiplications per second, which is limited by the framerate of the SWIR camera.

## 3. Experiments

*A. Characterization of the fabricated dot-product core*

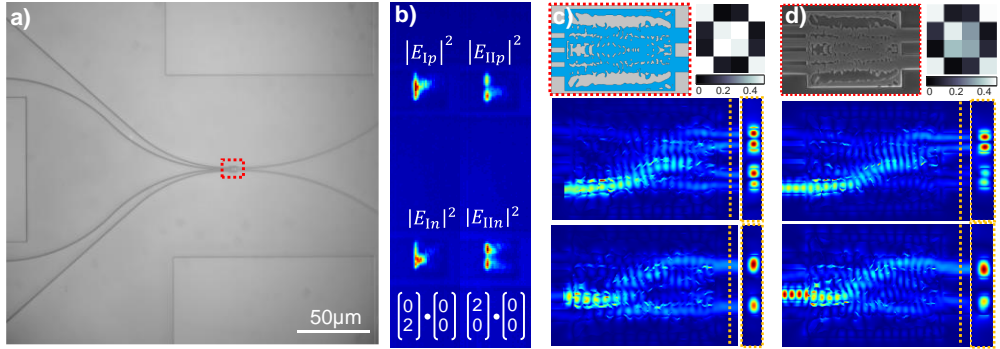

Figure 2. Characterization of the fabricated dot-product core. (a) Microscope image of the fabricated dot-product core under test. (b) Experimentally observed intensity profiles on the two output couplers when the inputs $a_1$ and $a_2$ were individually excited. (c) Structure of the ideal inversely designed dot-product core and simulated electrical field profiles within the core. (d) SEM image of the fabricated dot-product core and simulated electrical field profiles within the core based on the SEM image. The side views show the electrical field profiles at the location marked by the orange dashed line.

The fabricated core was edge-coupled to a fiber array probe that provided four modulated inputs, each driven by an independent MZM. Figure 2(a) shows a microscope image of the photonic core under our characterization setup. Figure 2(b) plots the intensity profiles at the output coupler when the first two single-mode input arms, $a_1$ and $a_2$, were individually activated in experiments. The intensity profiles match the target spatial profiles of the fundamental and second-order modes. To quantify the computing performance, we simulated the electromagnetic (EM) behavior of the ideal and fabricated core using Ansys Lumerical® based on the design and scanning electron microscope (SEM) image, respectively. TE fundamental modes were launched into each single-mode input waveguide, and the resulting field profiles at nominal operating wavelength 1570nm are shown in Figure 2(c) and (d). The orange boxes show the cross-section of the electrical field profiles marked by the dashed lines.

The transfer matrices of the ideal and fabricated designs can be calculated from the overlap integral [20] between the cross-sectional electrical fields and the two TE eigenmodes in the top and bottom arms. Both matrices share the same structure as Eq. (2). The ideal inversely designed core features a symmetric design with <10% crosstalk, as indicated by the off-diagonal elements. The power splitting ratios between the top and bottom arm are both approximately 46% vs. 54% for the fundamental and second-order TE modes. The fabricated core maintains the relative low crosstalk with a maximum of 13.5% in the off-diagonal elements. The power splitting ratios are 41% vs. 59%, 49% vs. 51% for fundamental and second-order TE modes, respectively. We experimentally compensated for the effects of uneven splitting in the dot-product results with the introduction of a calibration sequence at the beginning of each computing packet. Details of the compensation are discussed in the Supplementary Materials.

*B. General-purpose computing examples*

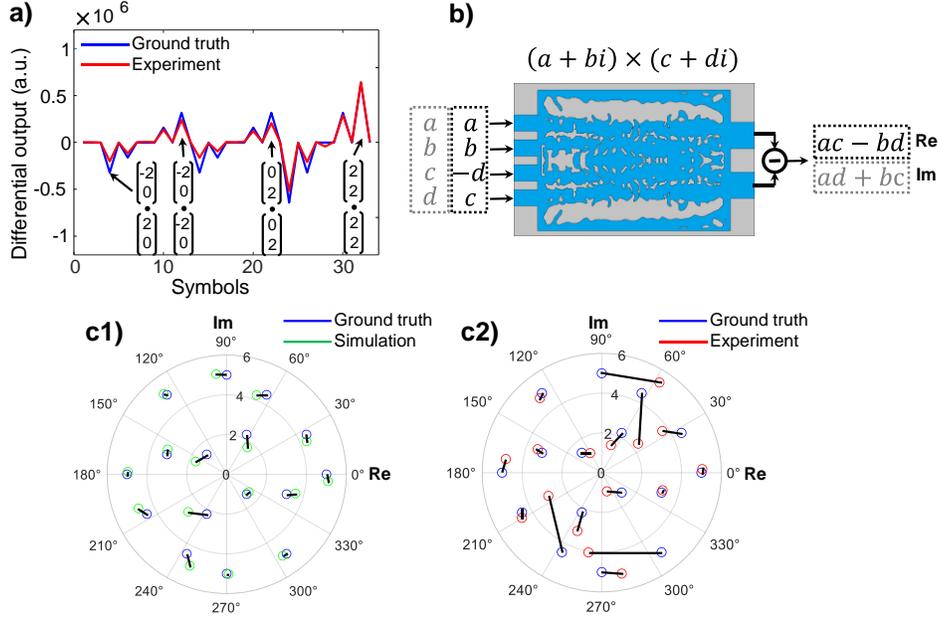

Figure 3. General-purpose computing examples as dot-products on the photonic core. (a) Dot-product calculation of a sequence of 16 2-element vectors. (b) Complex number multiplications encoded as two equivalent dot-products in time-division multiplexing. (c1, c2) Multiplication results between 16 complex numbers. Blue circles indicate ground truth results, green circles indicate simulated results from the ideal inversely designed core in (b), and red circles indicate experimental results calculated on the fabricated dot-product core.

The core supports dot-products between two 2-element vectors with fixed-point precision, enabling the deployment of general-purpose computing tasks such as complex number multiplication and optical flow calculation. To deploy general-purpose dot-product calculations on the photonic core, we calibrated the 4 MZMs to generate 5 signed linear analog levels representing the integers from -2 to 2. The intensity differences between two output couplers were proportionally mapped to the dot-products using the output from $(1,0) \cdot (1,0)^T$. Figure 3(a) plots a time-division multiplexing (TDM) sequence of 16 dot-products performed on the photonic core. We quantify the computing error with normalized mean square error (NMSE) between the ground truth $Y_{gt}$ and the experimental $Y_{exp}$ dot-products, defined in Eq. (7),

$$\text{NMSE} = \frac{\sum_{k=1}^{K} |Y_{k,exp} - Y_{k,gt}|^2}{\sum_{k=1}^{K} |Y_{k,gt}|^2}. \tag{7}$$

Here the summation is performed over all $K$ symbols in the sequence. The NMSE of all multiplications was 6.32%, offering sufficient dynamic range to represent signed integers from -8 to 8 in the results.

We first applied the photonic dot-product core to perform complex number multiplication, i.e., $(a + bi) \times (c + di)$. The real and imaginary parts of the result $((ac - bd) + (ad + bc)i)$ are split into two equivalent dot-products encoded in a TDM symbol sequence. 16 complex number pairs represented by a sequence of 32 dot-products were multiplied on the core. Figure 3(c) compares the products from the ideal and fabricated core with the ground truth on the complex plane. The designed dot-product core shows good agreement with ground truth and an NMSE of 4.0%, suggesting that the design can reach a dynamic range of signed 25 levels, or greater than signed 4-bit precision. The NMSE between the ground truth and experimental complex products is 15.9%, which is consistent with the simulation of a fabricated dot-product core. The computing error is primarily attributed to the fabrication deviation from the ideal design and the time-varying phase instability from the off-chip fiber inputs. The phase stability can be improved by switching to on-chip modulators. The fabrication deviation can be compensated with additional phase modulation on each input, which can be generated from integrated thermal optical phase shifters.

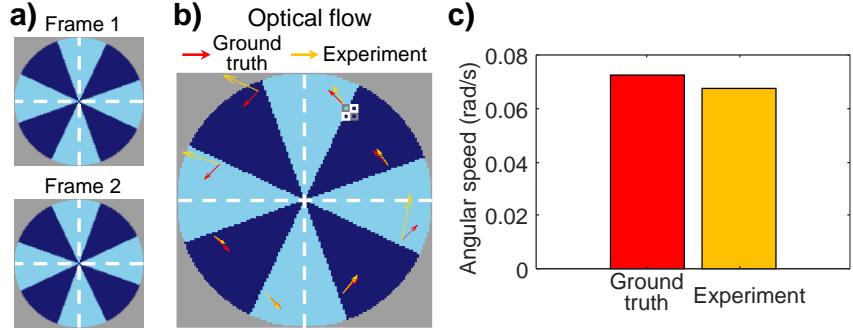

Figure 4. Optical flow calculation between two adjacent frames of a spinning wheel animation. (a) Two frames from the spinning wheel animation with 100ms interval (10 frames per second). (b) Optical flow vector of eight edge pixels. Red arrows indicate the ground truth of the flow vectors, and orange arrows indicate experimental results calculated on the photonic dot-product core. (c) Comparison of the calculated angular speed on the dot-product core with ground truth.

In addition, we have also demonstrated a computer vision task using the photonic dot-product core. Specifically, we use the device to calculate the optical flow in a visual scene to quantify the motion of the object. The real-time calculation of optical flow in a dynamic environment plays an important role in motion detection and object tracking of computer vision systems [21], [22]. Here we calculated the optical flow of selected edge pixels between two adjacent two-dimensional frames, $I_1(x,y)$ and $I_2(x,y)$, from a 10 frames-per-second spinning wheel animation on the dot-product core. The flow vector $(u,v)^T$ satisfies $(d_x, d_y) \cdot (u,v)^T = -d_t$, where $d_x$, $d_y$, and $d_t$ are the finite differences of the image $I_t(x,y)$ along $x$, $y$, and $t$ dimensions, respectively [23]. Due to the ambiguity in uniquely determining the pixelwise $(u,v)^T$, we expand the optical flow vector onto the diagonal pixels in a 2×2 window, as shown in Eq. (8),

$$\begin{bmatrix} d_{x11} & d_{y11} \\ d_{x22} & d_{y22} \end{bmatrix} \cdot \begin{bmatrix} u \\ v \end{bmatrix} = -\begin{bmatrix} d_{t11} \\ d_{t22} \end{bmatrix}. \tag{8}$$

Assuming uniform flow vectors in the 2×2 window, the calculations are broken down into two parts: 1) on the two pixels (marked in gray in Figure 4(b)) along the primary diagonals $d_{x11}$, $d_{x22}$, $d_{y11}$, $d_{y22}$, $d_{t11}$, $d_{t22}$; and 2) along the secondary diagonals $d_{x12}$, $d_{x21}$, $d_{y12}$, $d_{y21}$, $d_{t12}$, $d_{t21}$ (marked in white in Figure 4(b)). Results from primary and secondary diagonals are averaged to obtain the flow vector within the 2×2 window. Eq. (8) can be solved using Cramer's rule (Eq. (9)),

$$u = -\frac{\begin{vmatrix} d_{t11} & d_{y11} \\ d_{t22} & d_{y22} \end{vmatrix}}{\begin{vmatrix} d_{x11} & d_{y11} \\ d_{x22} & d_{y22} \end{vmatrix}}, v = -\frac{\begin{vmatrix} d_{x11} & d_{t11} \\ d_{x22} & d_{t22} \end{vmatrix}}{\begin{vmatrix} d_{x11} & d_{y11} \\ d_{x22} & d_{y22} \end{vmatrix}}. \tag{9}$$

Here, all the 2×2 determinants $\begin{vmatrix} a & b \\ c & d \end{vmatrix}$ are computed as their equivalent dot-products $ad - bc$. The flow vector within one 2×2 window requires 6 dot-product calculations encoded in a TDM sequence.

Figure 4(c) shows the optical flow vector within eight ($J = 8$) 2×2 windows on the edges of the spinning wheel. We quantify the error in flow vector calculation using the mean cosine similarity, defined in Eq. (10),

$$S_c = \frac{1}{J}\sum_{j=1}^{J} \frac{(u_{j,ideal}, v_{j,ideal}) \cdot (u_{j,exp}, v_{j,exp})}{\|(u_{j,ideal}, v_{j,ideal})\|\|(u_{j,exp}, v_{j,exp})\|}. \tag{10}$$

Here $(u_{j,ideal}, v_{j,ideal})$ and $(u_{j,exp}, v_{j,exp})$ represent the ideal flow vector and the one calculated on the fabricated dot-product core, respectively. The mean cosine similarity is 81.8%, suggesting that the flow vectors calculated on the photonic dot-product core have overall correct directions. The mean magnitude of the flow vectors reflects the angular speed of the wheel, which is 1.32 rad/s based on the calculated optical flow on the fabricated dot-product core. Compared with the ground truth 1.41 rad/s, the relative error of the angular speed calculation is 6.8%. This example illustrates that the fixed-point dot-product core can be used in conjunction with a tailored algorithm to extract features-of-interest in computer vision tasks.

## 4. Conclusion and discussions

In summary, we have demonstrated a compact, integrated photonic dot-product core from inverse design. The core utilizes spatial mode as the multiplexing dimension to perform arbitrary 2-element vector dot-products. We have shown the deployment of a general-purpose complex multiplier and an optical flow calculator on the dot-product core. The miniaturized footprint enables the large-scale integration of the core as part of the photonic primitives in electronic-photonic co-packaged parallel computing array. With on-chip modulators and multi-mode photodiodes [24], computing speed on the order of $10^9$ dot-products per second is supported by the modern giga-baud optoelectronics. By further integrating wavelength channels and spatial modes as super-dimensions in photonic matrix-/tensor-based processors [10], our strategy enables a computing throughput on the order of $10^3$ TOPS/mm$^2$, which is orders of magnitude higher than dedicated electronic vector/matrix accelerators [4], [25].

**Funding.** Office of Naval Research (N00014-20-1-2441), Army Research Office (W911NF2110321), National Science Foundation (1932858).

**Acknowledgments.** R.S. acknowledges the support by the Laboratory Directed Research and Development program at Sandia National Laboratories, a multi-mission laboratory managed and operated by National Technology and Engineering Solutions of Sandia, LLC, a wholly owned subsidiary of Honeywell International, Inc., for the U.S. Department of Energy's National Nuclear Security Administration under contract DE-NA-003525. This paper describes objective technical results and analysis. Any subjective views or opinions that might be expressed in the paper do not necessarily represent the views of the U.S. Department of Energy or the United States Government.

**Disclosures.** The authors declare no conflicts of interest.

**Data availability.** Data underlying the results presented in this paper are available in the main text and supplementary materials. Raw data may be obtained from the authors upon reasonable request.

**Supplemental document.** See Supplement 1 for supporting content.

## References

1. L. S. Blackford, A. Petitet, R. Pozo, K. Remington, R. C. Whaley, J. Demmel, J. Dongarra, I. Duff, S. Hammarling, and G. Henry, "An updated set of basic linear algebra subprograms (BLAS)," ACM Trans. Math. Softw. **28**, 135–151 (2002).
2. M. J. Flynn, "Some Computer Organizations and Their Effectiveness," IEEE Trans. Comput. **C–21**, 948–960 (1972).
3. H. Kaul, M. A. Anders, S. K. Mathew, S. K. Hsu, A. Agarwal, R. K. Krishnamurthy, and S. Borkar, "A 300 mV 494GOPS/W Reconfigurable Dual-Supply 4-Way SIMD Vector Processing Accelerator in 45 nm CMOS," IEEE J. Solid-State Circuits **45**, 95–102 (2010).
4. S. K. Hsu, A. Agarwal, M. A. Anders, S. K. Mathew, H. Kaul, F. Sheikh, and R. K. Krishnamurthy, "A 280 mV-to-1.1 V 256b Reconfigurable SIMD Vector Permutation Engine With 2-Dimensional Shuffle in 22 nm Tri-Gate CMOS," IEEE J. Solid-State Circuits **48**, 118–127 (2013).

5. Y. Shen, N. C. Harris, S. Skirlo, M. Prabhu, T. Baehr-Jones, M. Hochberg, X. Sun, S. Zhao, H. Larochelle, D. Englund, and M. Soljačić, "Deep learning with coherent nanophotonic circuits," Nat. Photonics **11**, 441–446 (2017).
6. R. Hamerly, L. Bernstein, A. Sludds, M. Soljačić, and D. Englund, "Large-Scale Optical Neural Networks Based on Photoelectric Multiplication," Phys. Rev. X **9**, 021032 (2019).
7. B. J. Shastri, A. N. Tait, T. Ferreira de Lima, W. H. P. Pernice, H. Bhaskaran, C. D. Wright, and P. R. Prucnal, "Photonics for artificial intelligence and neuromorphic computing," Nat. Photonics **15**, 102–114 (2021).
8. K. Kikuchi, "Fundamentals of Coherent Optical Fiber Communications," J. Light. Technol. **34**, 157–179 (2016).
9. B. Murmann, "ADC performance survey 1997-2022," http//www. stanford. edu/~ murmann/adcsurvey. html (2022).
10. A. Fardoost, F. G. Vanani, Z. Zhu, C. Doerr, S. Pang, and G. Li, "A High-Speed Photonic Tensor Accelerator," in *2022 IEEE Photonics Conference (IPC)* (IEEE, 2022), pp. 1–2.
11. M. A. Nahmias, T. F. de Lima, A. N. Tait, H.-T. Peng, B. J. Shastri, and P. R. Prucnal, "Photonic Multiply-Accumulate Operations for Neural Networks," IEEE J. Sel. Top. Quantum Electron. **26**, 1–18 (2020).
12. A. N. Tait, M. A. Nahmias, B. J. Shastri, and P. R. Prucnal, "Broadcast and weight: An integrated network for scalable photonic spike processing," J. Light. Technol. **32**, 3427–3439 (2014).
13. T. F. de Lima, H.-T. Peng, A. N. Tait, M. A. Nahmias, H. B. Miller, B. J. Shastri, and P. R. Prucnal, "Machine Learning With Neuromorphic Photonics," J. Light. Technol. **37**, 1515–1534 (2019).
14. X. Wu, C. Huang, K. Xu, C. Shu, and H. K. Tsang, "Mode-Division Multiplexing for Silicon Photonic Network-on-Chip," J. Light. Technol. **35**, 3223–3228 (2017).
15. K. Y. Yang, C. Shirpurkar, A. D. White, J. Zang, L. Chang, F. Ashtiani, M. A. Guidry, D. M. Lukin, S. V. Pericherla, J. Yang, H. Kwon, J. Lu, G. H. Ahn, K. Van Gasse, Y. Jin, S.-P. Yu, T. C. Briles, J. R. Stone, D. R. Carlson, H. Song, K. Zou, H. Zhou, K. Pang, H. Hao, L. Trask, M. Li, A. Netherton, L. Rechtman, J. S. Stone, J. L. Skarda, L. Su, D. Vercruysse, J.-P. W. MacLean, S. Aghaeimeibodi, M.-J. Li, D. A. B. Miller, D. M. Marom, A. E. Willner, J. E. Bowers, S. B. Papp, P. J. Delfyett, F. Aflatouni, and J. Vučković, "Multi-dimensional data transmission using inverse-designed silicon photonics and microcombs," Nat. Commun. **13**, 7862 (2022).
16. D. Dai, J. Wang, and Y. Shi, "Silicon mode (de)multiplexer enabling high capacity photonic networks-on-chip with a single-wavelength-carrier light," Opt. Lett. **38**, 1422 (2013).
17. G. Zhang, H. R. Mojaver, A. Das, and O. Liboiron-Ladouceur, "Mode insensitive switch for on-chip interconnect mode division multiplexing systems," Opt. Lett. **45**, 811 (2020).
18. H. Shiran, G. Zhang, and O. Liboiron-Ladouceur, "Dual-mode broadband compact 2 × 2 optical power splitter using sub-wavelength metamaterial structures," Opt. Express **29**, 23864 (2021).
19. Y. Zhang, M. A. Al-Mumin, H. Liu, C. Xu, L. Zhang, P. L. LiKamWa, and G. Li, "An Integrated Few-Mode Power Splitter Based on Multimode Interference," J. Light. Technol. **37**, 3000–3008 (2019).

# Supplementary materials: A mode-multiplexed photonic integrated vector dot-product core from inverse design

Zheyuan Zhu[1], Raktim Sarma[2], Seth Smith-Dryden[1], Guifang Li[1], and Shuo S. Pang[1]

[1]CREOL, The College of Optics and Photonics, University of Central Florida, 4304 Scorpius St. Orlando, FL 32816-2700

[2]Center for Integrated Nanotechnologies, Sandia National Laboratories, 1515 Eubank Blvd SE, Albuquerque, NM 87185

## S1. Characterization of fabricated dot-product core

The behavior of the ideal and fabricated dot-product core can be characterized by the transfer matrix, $S_t$, which describes the fraction of the electromagnetic (EM) field coupled into the individual target modes, $\psi_I$ and $\psi_{II}$, in the few-mode output waveguides. Based on the EM field $E_j$ within the device (Fig. 2(c) and (d) in the main text) under the excitation from the input arm $j$, we calculate $S_t$ as in Eq. (S1)

$$S_t[i,j] = E_j^\dagger \psi_i. \tag{S1}$$

A more straightforward metric to quantify the insertion loss and crosstalk is the crosstalk matrix, $M_X$, whose elements are the overlap integral between the corresponding columns in the transfer matrices of the target ($S_i$) and designed (or fabricated) device ($S_t$),

$$M_X[i,j] = S_t[:,i]^* \cdot S_i[:,j]. \tag{S2}$$

Here $[:,i]$ extracts the $i$-th column vector from the matrix. The insertion loss (IL) and crosstalk (XT) can both be derived from $M_X$ respectively as in Eq. (S3) [1],

$$\text{IL(dB)} = -10\log_{10}(\max \text{ eigenvalue of } M_X); \tag{S3}$$
$$\text{XT(dB)} = -10\log_{10}\left(\frac{\text{power in the diagonals of } M_X}{\text{power in the off}-\text{diagonals of } M_X}\right).$$

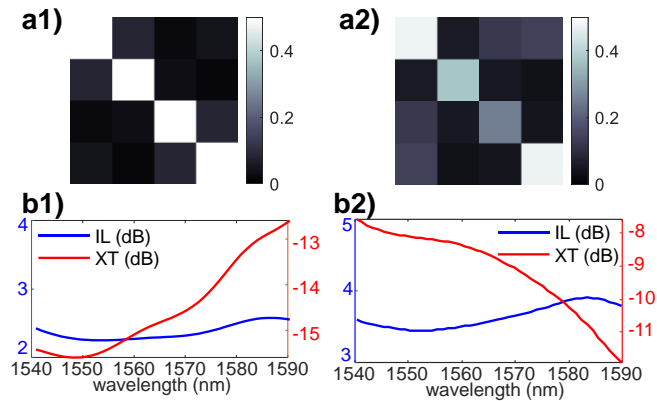

Figure S1. Characterization of ideal (a1, b1) and fabricated (a2, b2) dot-product core. (a). Crosstalk matrix $M_X$ of the core. (b) Insertion loss and crosstalk (in dB) as a function of wavelength.

The crosstalk matrices at normal operating wavelength 1570nm of both ideal and fabricated dot-product core design are shown in Figure S1(a). Figure S1(b) plots the insertion loss and crosstalk of the designed and fabricated core as a function of wavelength. The ideal dot-product core design features a consistent 2.3dB insertion loss and a crosstalk of <-13dB (<5%) across the wavelength range of 1540nm to 1590nm. The fabricated core maintains consistent insertion loss and crosstalk within the wavelength range 1550 to 1580nm, suggesting broadband performance that supports wavelength multiplexed inputs. Despite the uneven splitting of the input fields into the upper and lower arms, the crosstalk between the two spatial modes in the output waveguides is -9.06dB, or 12.4%. The low crosstalk allows us to empirically correct most of the computing errors, as described in S2.

## S2. Calibration and computing error of dot-product core

Given the transfer matrix, $S_t$, of the fabricated dot-product core, it is possible to compensate the four inputs to account for the uneven splitting and/or crosstalk due to fabrication imperfections. The compensation mixes the four inputs according to the weights in a matrix, $C$, before sending to the dot-product core. The choice of $C$ must minimize the crosstalk and equalize the amplitudes in the two output arms for both spatial modes. Theoretically, $C$ can be calculated according to Eq. (S4),

$$S_t \cdot C = \frac{1}{\sqrt{2}} \begin{bmatrix} 1 & 0 & 0 & i \\ 0 & 1 & i & 0 \\ 0 & i & 1 & 0 \\ i & 0 & 0 & 1 \end{bmatrix}. \tag{S4}$$

Here the right-hand side denotes the transfer matrix of an ideal dot-product core in Eq. (2). Figure S2 shows the transfer matrix of the fabricated core and the corresponding compensation matrix $\tilde{C}_{full}$ for pre-mixing the four inputs. However, using the full transfer matrix for compensation involves the multiplication of a 4×4 complex matrix on top of the desired inputs, giving rise to 16 additional digital MAC operations.

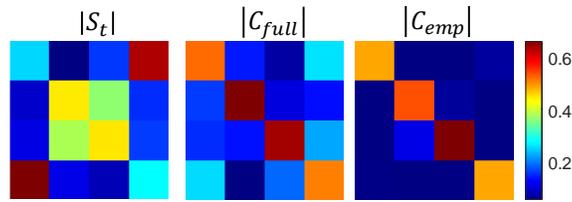

Figure S2. Transfer matrix of the fabricated dot-product core, $S_t$, and the corresponding compensation matrices $C_{full}$ and $C_{emp}$. Only the magnitudes of the matrix elements are shown.

Because of the low crosstalk among the four output field profiles, we can also empirically treat two spatial modes independently and ignore the off-diagonal elements in the transfer matrix when correcting the dot-products in experiments. This is equivalent to constraining all the off-diagonal elements in $S_t$ and its associated compensation matrix $\tilde{C}_{emp}$ to zero. The approximated transfer matrix of the fabricated core is shown in Eq. (S5),

$$\begin{bmatrix} E_{IIp} \\ E_{Ip} \\ E_{In} \\ E_{IIn} \end{bmatrix} = \begin{bmatrix} S_{II11} & 0 & 0 & S_{II12} \\ 0 & S_{I11} & S_{I12} & 0 \\ 0 & S_{I21} & S_{I22} & 0 \\ S_{II21} & 0 & 0 & S_{II22} \end{bmatrix} \begin{bmatrix} a_1 \\ a_2 \\ b_2 \\ b_1 \end{bmatrix}. \quad (S5)$$

Based on Eq. (S5), the empirical compensation can be formulated as a complex matrix (Eq. (S4)) in which only two elements in each row are non-zero. It is worth noting that experimentally implementing a complex compensation matrix $\tilde{C}_{emp}$ entails generating accurate phase modulations on individual inputs, which are hindered by the intrinsic time-varying phase on the four off-chip inputs.

To perform empirical compensation based solely on the amplitude scaling, we observe the intensity difference between the two output couplers from Eq. (S5),

$$\begin{aligned} I_{diff} = &|a_2|^2(|S_{I11}|^2 - |S_{I21}|^2) + |b_2|^2(|S_{I12}|^2 - |S_{I22}|^2) + 2\text{Re}[a_2 b_2^*(S_{I11}S_{I12}^* - S_{I21}S_{I22}^*)] \\ &+ |a_1|^2(|S_{II11}|^2 - |S_{II21}|^2) + |b_1|^2(|S_{II12}|^2 - |S_{II22}|^2) \\ &+ 2\text{Re}[a_1 b_1^*(S_{II11}S_{II12}^* - S_{II21}S_{II22}^*)]. \end{aligned} \quad (S6)$$

Eq. (S6) indicates that to produce the result proportional to the dot-product $a_1 b_1 + a_2 b_2$, the residue terms $|a_2|^2(|S_{I11}|^2 - |S_{I21}|^2) + |b_2|^2(|S_{I12}|^2 - |S_{I22}|^2) + |a_1|^2(|S_{II11}|^2 - |S_{II21}|^2) + |b_1|^2(|S_{II12}|^2 - |S_{II22}|^2)$ arising from the uneven splitting can be corrected by removing the differential outputs when only one of the four inputs is excited. The amplitude scaling factors $\text{Re}(S_{I11}S_{I12}^* - S_{I21}S_{I22}^*)$ and $\text{Re}(S_{II11}S_{II12}^* - S_{II21}S_{II22}^*)$ can be merged into the calibrated amplitude of $b_1$ and $b_2$.

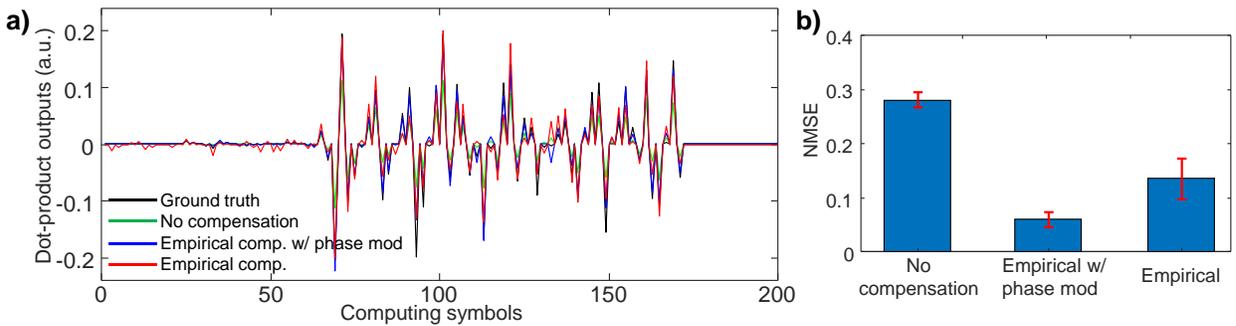

Figure S3. Comparison between different compensation methods. (a) TDM dot-product output sequence before and after compensation using the full transfer matrix $S_t$ and the empirical method with/without phase modulation. (b) Comparison between the NMSE of the raw and compensated dot-products.

Based on the transfer matrix $S_t$ of the fabricated core, Figure S3 simulates the computing errors before and after compensation with ten 40-symbol random sequences of dot-products deployed on the fabricated dot-product core. Both compensation methods are effective in reducing computing errors by at least two folds, giving a nominal NMSE of ~15% after empirical compensation. The empirical compensation with amplitude scaling only involves remapping the amplitude of the individual inputs, which can be efficiently implemented as an input look-up table on a microcontroller and is thus computationally efficient. In experiments, we appended all the single-input excitations as a calibration header and use them to remove residue terms from the computing symbols, as shown in Figure S4.

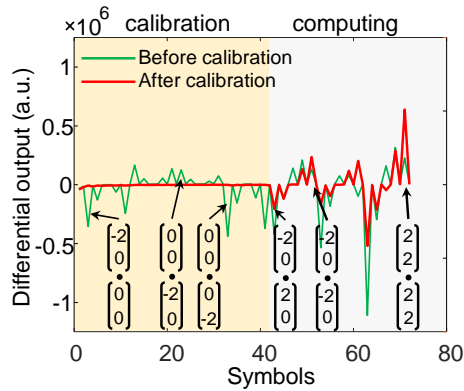

Figure S4. Comparison between the dot-products before and after calibration in experiments.

The NMSE after empirical compensation presented here represents a theoretical lower bound in the computing error. In actual experiments, the time-varying phase on the four off-chip modulated inputs cannot be measured and compensated. As a result, the experimental computing error could be higher than the lower bound. We envision that with fully integrated optical paths on chip, including the use of on-chip modulators [2] and few-mode photodiodes [3], the time-varying phase could be resolved.

# References

[1]   N. K. Fontaine, R. Ryf, H. Chen, D. Neilson, and J. Carpenter, "Design of High Order Mode-Multiplexers using Multiplane Light Conversion," Sep. 2017, doi: 10.1109/ECOC.2017.8346129.

[2]   J. Sun, R. Kumar, M. Sakib, J. B. Driscoll, H. Jayatilleka, and H. Rong, "A 128 Gb/s PAM4 Silicon Microring Modulator With Integrated Thermo-Optic Resonance Tuning," *J. Light. Technol.*, vol. 37, no. 1, pp. 110–115, Jan. 2019, doi: 10.1109/JLT.2018.2878327.

[3]   L. Wu, D. Lv, N. Zhao, R. Wang, and A. Wu, "Research on Germanium Photodetector with Multi-Mode Waveguide Input," *Photonics*, vol. 10, no. 4, p. 455, Apr. 2023, doi: 10.3390/photonics10040455.